\documentclass[11pt]{article}
\usepackage{latexsym}

\usepackage{latexsym}
\usepackage{bm}

\def\tr{{\rm tr}}
\def\ket#1{\mid~\!\!\!{#1}~\!\!\rangle}
\def\bra#1{\langle~\!\!{#1}~\!\!\!\mid}
\def\IF{if and only if }
\def\MI{measuring instrument }

\def\QM{quantum mechanics }
\def\qm{quantum mechanics}

\def\CC{calibration condition }
\def\cc{calibration condition}
\def\PRC{probability reproducibility condition }
\def\prc{probability reproducibility condition}

\def\ON{orthonormal }

\def\${\enskip$}
\def\M{measurement }
\def\m{measurement}
\def\Q{quantum }

\begin{document}

\begin{center}
{\bf\noindent \large DERIVATION OF QUANTUM PROBABILITY FROM MEASUREMENT}
\vspace{0.3cm}

{ \large\noindent FEDOR HERBUT*}\\

{\rm \noindent Serbian Academy of
Sciences and Arts, Knez Mihajlova 35,
11000 Belgrade, Serbia}
\vspace{0.3cm}

\date{\today}

\end{center}

\noindent
To begin with, it is pointed out that the form of the \Q probability formula originates in the very initial state of the object system as seen when the state is expanded with the eigen-projectors of the measured observable. Making use of the probability reproducibility condition, which is a key concept in unitary \M theory, one obtains the relevant coherent distribution of the complete-\M results in the final unitary-\M state in agreement with the mentioned probability formula. Treating the transition from  the final unitary, or pre\m , state, where all possible results are present, to one complete-\M result sketchily in the usual way, the      well-known probability formula is derived. In conclusion it is pointed out that the entire argument is only formal unless one makes it physical assuming that the \Q probability law is valid in the extreme case of probability-one (certain) events (projectors).

\vspace{0.5cm}

\normalsize \rm

\section{Introduction}

\noindent
Probability has no physical meaning if \M is not taken into account. Hence, the physically most appropriate way to derive probability is to do it in the framework of \M theory.  I have demonstrated advantages of such a procedure within Zurek's way to derive probability from 'envariance' (invariance due to entanglement) \cite{FHenvar2}.\\

Complete \M that will be utilized for our derivation consists of two parts: Relevant parts of unitary \M theory (also called pre\M theory or \M theory short of collapse) and a sketchy phenomenological idea of collapse.

Unitary \M theory will be along the lines of former work \cite{FHijqi1}, which allowed for redundant entanglement. The basic concepts of this approach,
which was based on an unpublished but detailed and systematic

{\footnotesize \rm \noindent
\rule[0mm]{4.62cm}{0.5mm}

\noindent *e-mail: fedorh@sanu.ac.rs}

\noindent
review \cite{FHarxiv}, will be outlined now.

The observables treated in this article are confined to discrete ones, i. e., to ordinary (as opposed to generalized) observables
that do not have a continuous part in their spectrum. The object of \M is denoted by A, and the \MI by B. The measured observable \$O\$ is given in its unique spectral form (in which, by definition, there is no repetition in the eigenvalues \$\{o_k:\forall k\}\$:
$$O_A=\sum_ko_kE_A^k.\eqno{(1a)}$$ The eigen-projectors \$\{E_A^k:\forall k\}\$ satisfy the completeness relation $$ \sum_kE_A^k=I_A, \eqno{(1b)}$$ where \$I_A\$ is the identity operator  in the state space of the object subsystem.

The \MI has a suitable initial state \$\ket{\phi}_B^i\$ and a so-called pointer observable, which in its unique spectral form reads $$P_B=\sum_kp_kF_B^k.\eqno{(2a)}$$ There is also the completeness relation $$\sum_kF_B^k=I_B.\eqno{(2b)}$$
The coindexing is due to a one-to-one relation between the spectral form of the measured observable and that of the \MI with the physical meaning that the result \$o_k\$ (or equivalently the occurrence of \$E_A^k\$) is noted by the \MI by the occurrence of the so-called pointer position \$F_B^k$.

Finally, there is the unitary operator \$U_{AB}\$ that includes the object-measuring-instrument interaction and transforms the initial state \$\ket{\phi}_A^i\ket{\phi}_B^i\$ of object+\MI into the final state $$\ket{\Phi}_{AB}^f\equiv U_{AB}\Big(\ket{\phi}_A^i\ket{\phi}_B^i\Big). \eqno{(3)}$$\\

Exact \M (as opposed to approximate \m ) in its general form (as opposed to the particular case of nondemolition \M or the evn more special case of ideal \m , cf \cite{FHarxiv}) is defined by the {\bf \CC } for discrete observables (cf \cite{BLM}):
$$\bra{\phi}_A^iE_A^{\bar k}\ket{\phi}^i=1\quad
\Rightarrow\quad \bra{\Phi}_{AB}^fF_B^{\bar k}
\ket{\Phi}_{AB}^f=1,\eqno{(4)}$$, which can be equivalently rewritten in the more practical form:
$$E_A^{\bar k}\ket{\phi}_A^i=\ket{\phi}_A^i\quad
|\Rightarrow\quad F_B^{\bar k}
\ket{\Phi}_{AB}^f=\ket{\Phi}_{AB}^f.\eqno{(5)}$$
(The equivalence of (4) and (5) is easily proved.).

It was shown in previous work \cite{FHijqi1} that the \CC is equivalent to the {\bf dynamical condition}:
$$\forall k:\quad F_B^KU_{AB}\Big(\ket{\phi}_A^i\ket{\phi}_B^i\Big)=
U_{AB}E_A^k\Big(\ket{\phi}_A^i\ket{\phi}_B^i\Big).
\eqno{(6})$$ (For the reader's convenience the exposition is in this article self-contained. To this purpose, the proof of the claimed equivalence is reproduced in Appendix B.)\\

\section{Role of the Probability Reproducibility Condition}
\noindent

For our purpose, let it be pointed out that an arbitrary state \$\ket{\phi}_A^i\$ of the object has, on account of the completeness relation (1b), the following decomposition:
$$\forall \ket{\phi}_A^i:\quad \ket{\phi}_A^i=
\sum_k||E_A^k\ket{\phi}_A^i||\times (E_A^k\ket{\phi}_A^i\Big/
||E_A^k\ket{\phi}_A^i||) \eqno{(7)}$$ where it is understood that if the first factor in a term is zero, that the entire term is zero though the second factor does not make sense.

Further, due to idempotency of the projectors \$\{E_A^k:\forall k\}]$, $$\forall \ket{\phi}_A^i:\quad
||E_A^k\ket{\phi}_A^i||=(\bra{\phi}_A^iE_A^k \ket{\phi}_A^i)^{1/2}.\eqno{(8)}$$
In our derivation this is, excuse the pun, where the Born rule is borne.

Incidentally, the strict form of the Born rule, the most used expression for pure states and the trace rule are all mutually equivalent forms of the probability law in \QM (as proved in Appendix A). We are going to derive it.

A key role is played in unitary \M theory by the so-called {\bf \prc }: $$\forall \ket{\phi}_A^i:\quad\bra{\Phi}_{AB}^fF_B^k \ket{\Phi}_{AB}^f=\bra{\phi}_A^iE_A^k \ket{\phi}_A^i.\eqno{(9)}$$
It was shown in previous work \cite{FHijqi1} how the \PRC follows from the \cc . (The proof is reproduced in Appendix C.)\\

Now we can derive the relevant decomposition of the final state. Making use of the completeness relation (2b) and the idempotency of the projectors \$\{F_B^k:\forall k\}\$, one can write: $$\ket{\Phi}_{AB}^f=\sum_kF_B^k\ket{\Phi}_{AB}^f=
\sum_k||F_B^k\ket{\Phi}_{AB}^f||\times
F_B^k\ket{\Phi}_{AB}^f\Big/
||F_B^k\ket{\Phi}_{AB}^f||=$$   $$
\sum_k\Big(\bra{\Phi}_{AB}^fF_B^k
\ket{\Phi}_{AB}^f\Big)^{1/2}\times
F_B^k\ket{\Phi}_{AB}^f\Big/
||F_B^k\ket{\Phi}_{AB}^f||.$$ Finally, the \PRC (9)
gives $$\ket{\Phi}_{AB}^f=\sum_k
\Big(\bra{\phi}_A^i)E_A^k
 \ket{\phi}_A^i)\Big)^{1/2}\times
 F_B^k\ket{\Phi}_{AB}^f\Big/
||F_B^k\ket{\Phi}_{AB}^f||.\eqno{(10)}$$\\

\section{The Final Steps}
\noindent

In the final steps we have to leave the unitary final state \$\ket{\Phi}_{AB}^f\$ (cf (3)) and reach the result of {\bf complete \M } to which corresponds one value of \$k\$ for an individual object - the so-called collapse of the unitary final state. Unitary \QM cannot do this
(unless we accept the many-worlds interpretation, which we will not do now).

Peres in his book \cite{Peres} (the last chapter there) speaks of dequantization when it comes to complete \m . Accepting the Copenhagen interpretation of \qm , his dequantization  consists in the assumption that the pointer-position projectors \$\{F_B^k:\forall k\}\$ represent classical events.
Viewing the completeness relation (2b) classically only one of the mutually excluding events ca happen
Thus the complete \M results are obtained.

Bell criticized collapse \cite{Bell} viewing it entirely within \qm . The \Q entity that has to collapse, written as a density operator is: $$\ket{\Phi}_{AB}^f\bra{\Phi}_{AB}^f=\sum_k\sum_{k'}
\Big(\bra{\phi}_A^i)E_A^k
 \ket{\phi}_A^i)\Big)^{1/2}\Big(\bra{\phi}_A^i)E_A^{k'}
 \ket{\phi}_A^i)\Big)^{1/2}\times$$  $$\Big(
 F_B^k\ket{\Phi}_{AB}^f\Big/
||F_B^k\ket{\Phi}_{AB}^f||\Big)
\Big(\bra{\Phi}_{AB}^fF_B^{k'}\Big/
||F_B^{k'}\ket{\Phi}_{AB}^f||\Big).\eqno{(11)}$$

One should note that in (11), besides the diagonal (k=k') terms also the off-diagonal ($k\not= k'$) terms are non-zero (each for some initial state). The latter express {\bf coherence}. They must be deleted in collapse. Thus, the first step is replacing the LHS(11) by $$\rho_{AB}\equiv
\sum_k\bra{\phi}_A^i)E_A^k
 \ket{\phi}_A^i\times$$  $$
 F_B^k\ket{\Phi}_{AB}^f\Big/
||F_B^k\ket{\Phi}_{AB}^f||
\bra{\Phi}_{AB}^fF_B^k\Big/
||F_B^k\ket{\Phi}_{AB}^f||\Big).\eqno{(12)}$$
Bell called \$\rho_{AB}\$ the "butchered state".

In spite of butchering the coherence in \$\ket{\Phi}_{AB}^f\bra{\Phi}_{AB}^f\$ one would expect that \$\rho_{AB}\$ still represents the state of individual \Q systems as the former state did. But, in the second step of collapse, one assumes that \$\rho_{AB}\$ given by (12) describes the state of an ensemble in which the states of the individual systems are described by the pure states in the terms in (12). So that (12) is assumed to represent a {\bf mixture} with the
{\bf statistical weights} $$\forall k:\qquad w_k\equiv\bra{\phi}_A^i)E_A^k
 \ket{\phi}_A^i.\eqno{(13)}$$ Bell called this step "replacing "or" by "and"".

The final and for our purpose the most important step is assuming that the {\bf probability}  of obtaining the result \$F_B^k\ket{\Phi}_{AB}^f\Big/
||F_B^k\ket{\Phi}_{AB}^f||\$ in complete \M {\bf equals the statistical weight} \$w_k\$ given by (13). This ends our argument of deriving the \Q probability law from general \m , at least its formal part. It is physically completed in concluding remark C in section 5.\\

One should note that the steps that we have made use of in this section are actually phenomenological, i. e., we know from experience that these steps are made in complete \m .\\

\section{The Mixed Initial State Case}
\noindent

Now we assume that the initial state of the object
system is a general state. (Our interest lies, of course, in mixed states because we have already dealt with the pure states.) The method called {\bf purification} will be applied to reduce general states to pure states.

We denote the object system by \$A_1\$. Let \$A_2\$ be another system, which will play only a formal role.

Let \$\rho_{A_1}^i=\sum_ir_i\ket{i}_{A_1} \bra{i}_{A_1}\$ be a decomposition of the given initial state of the object system \$A_1\$ into its
positive-eigenvalue norm-one eigenvectors. Further, let \$\{\ket{i}_{A_2}:\forall i\}\$ be an arbitrary \ON set of vectors in the state space of \$A_2\$. We define
$$\forall\enskip \rho_{A_1}^i:\quad \ket{\phi}_{A_1A_2}^i
\equiv\sum_ir_i\ket{i}_{A_1}\ket{i}_{A_2}.\eqno{(14a)}$$
The essential property of this composite-system pure state, which characterizes purification, is that
$$\tr_{A_2}\Big(\ket{\phi}_{A_1A_2}^i
\bra{\phi}_{A_1A_2}^i\Big)=\rho_{A_1}^i,\eqno{(14b)}$$
\$\rho_{A_1}^i\$ being the initial state of the object subsystem that we started with.\\

Let the measured observable be \$O_{A_1}=\sum_ko_k
E_{A_1}^k\$, and let the \MI be subsystem B with the initial state \$\ket{\phi}_B^i\$ and the pointer observable \$P_B=\sum_kp_kF_B^k\$ as before. Then, as proved in the preceding sections, the probability to obtain in complete \M of \$O_{A_1}\$ the state \$F_B^k\ket{\Phi}_{A_1A_2B}
\Big/||F_B^k\ket{\Phi}_{A_1A_2B}||
\$ is: $$\bra{\phi}_{A_1A_2}^iE_{A_1}^k
\ket{\phi}_{A_1A_2}^i.\eqno{(15)}$$

Now it is time for depurification, i. e., to rid ourselves of the passive subsystem \$A_2\$. The expectation value (15) is standardly rewritten in terms of its subsystem state operator ( reduced density operator) as:
$$\tr\Big(\rho_{A_1}^iE_{A_1}^k\Big).\eqno{(16)}$$
This is the final result.\\

\section{Concluding Remarks}
\noindent

{\bf A)} We have seen in relation (10) that the final state of unitary \M theory is a state in which all possible result are contained. In order to reach the final state of complete \M the two steps described sketchily in section 3 are unavoidable: one must terminate the coherence in (10) (the "butchering" following Bell), and then the drastic change that the butchered state \$\rho_{AB}\$ is not valid for individual systems, only for an ensemble of such, where the terms in (12) apply to the individual systems making up the ensemble (Bell's "replacing "and" by "or"").

The fact that derivation of a final state of complete \M is considered to be impossible in unitary \QM is known as the paradox of \Q \m . (Though the derivation is possible in the many-worlds interpretation of \qm , which is not universally accepted.)\\

{\bf B)} One might think of complete \M that does not end in the state \$F_B^k\ket{\Phi}_{AB}^f\Big/
||F_B^k\ket{\Phi}_{AB}^f||\$. This may be the case, e. g., if one has over\M \cite{FHOverMeas}. But then one deals with a different probability formula. The one derived in this study, which is the standard one (cf Appendix A) is better understood by the following explanation.

Utilizing (3), one can rewrite the dynamical condition (6) as follows $$\forall\ket{\phi}_A^\mathbf{i},\forall k:\enskip\ket{\phi}_A^\mathbf{i}\enskip\rightarrow\enskip
U_{AB}\Big(E_A^k\ket{\phi}_A^\mathbf{i}\ket{\phi}_B^\mathbf{i}\Big)=
F_B^k\ket{\Phi}_{AB}^\mathbf{f}.\eqno{(17)}$$
One can see that {\bf each initial term} \$E_A^k\ket{\phi}_A^\mathbf{i}\$ (cf relation (7)) {\bf evolves} (applying to it \$U_{AB}(\dots\otimes\ket{\phi}_B^\mathbf{i})\$)
{\bf separately}, i. e., independently of the rest of the terms, into the corresponding final term \$F_B^k\ket{\Phi}_{AB}\$. Thus, in  unitary \M we have a set of {\bf complete-\M branches, each evolving independently of each other, but tied up into a whole by coherence}.

Thus, we actually consider one {\bf entire} branch branch in seeking to reach the corresponding complete-\M state. We begin with a definite   eigenvalue state \$E_A^k\ket{\phi}_A^i\Big/||E_A^k\ket{\phi}_A^i||\$ with \$||E_A^k\ket{\phi}_A^i||=\Big(\bra{\phi}_A^iE_A^k
\ket{\phi}_A^i\Big)^{1/2}\$ (where the Born rule begins, as stated -  cf relation (8) and beneath it).

In over\M we would not start with the entire branch \$k\$. One would have \$E_A^k=\sum_{\bar k}E_A^{\bar k}\$ and one would endeavor to reach the complete-\M state corresponding to a fixed
\$\bar k\$ value. In the end, one would then derive \$\bra{\phi}_A^i\bar E_A^{\bar k}\ket{\phi}_A^i\$.\\

{\bf C)} The entire derivation in sections 2 and 3 is algebraic and formal. We must put in a suitable physical assumption at the beginning, so that we obtain a physically meaningful result at the end.

Since we have made essential use of the dynamical condition (6), and it is equivalent to the \CC (5),
it is the latter that must be given physical meaning. To do this the idea of a (statistically) sharp value  must be expressible as \$\bra{\phi}_A^iE_A^k
\ket{\phi}_A^i=1\$. Since the latter is equivalent to \$E_A^k\ket{\phi}_A^i=1\times \ket{\phi}_A^i\$, we must {\bf assume} that if an event (projector) in a pure state has the eigenvalue one, then the event is {\bf certain in this state.}

Thus, assuming the physical validity of the probability formula that is to be derived in the special extreme case, we obtain the physically meaningful final formula for all cases (13).

I have read somewhere that you cannot derive probability unless you put in something of probability. It is certainly valid for our derivation. Incidentally, a completely different derivation of the \Q probability law \cite{Farhi}
started with the same physical assumption.\\

{\bf Appendix A. Equivalent forms of the \Q probability law}\\
\noindent
Let P denote a projector and let \$\ket{\psi}\$ and \$\ket{\phi}\$ denote norm-one vectors. The following three probability expressions are equivalent:
$$\bra{\psi}P\ket{\psi}\enskip (1)\quad\Leftrightarrow\quad
|\bra{\psi}\ket{\phi}|^2\enskip (2) \quad\Leftrightarrow\quad
\tr(P\ket{\psi}\bra{\psi})\enskip (3).\eqno{(A.1)}$$

Expression (2) is the "Born rule" (in the strict sense), and expression (3) is called the "trace rule".\\

{\it Proof}. We assume that \$P=
\ket{\phi}\bra{\phi}\$. Then expression (1) becomes expression (2) as one can see using the Dirac rules.

Let \$P=\sum_k\ket{\phi,k}\bra{\phi,k}\$ be a complete orthogonal decomposition of \$P\$. Let us further  assume that the probability of an orthogonal sum (disjoint events) is sum of the probabilities of the terms. Then $$\sum_k|\bra{\psi}\ket{\phi,k}|^2 =
\bra{\psi}\Big(\sum_k\ket{\phi,k}\bra{\phi,k}
\Big)\ket{\psi}=
\bra{\psi}P\ket{\psi}.\eqno{(A.2)}$$ The first equivalence is proved.

Having in mind evaluation of the trace in a basis in which \$\ket{\psi}\$ is one of the basis vectors, one can see that $$\bra{\psi}P\ket{\psi}=
\tr(P\ket{\psi}\bra{\psi}).\eqno{(A.3)}$$ This proves the equivalence of (1) with (3). The second equivalence in (A.1) is then a consequence of transitivity of equivalences.\hfill $\Box$\\

{\bf Appendix B. Proof of the dynamical condition }\\
\noindent
We now express and prove the {\bf dynamical condition}, valid for general \m , and being {\bf equivalent to the \cc }.

The {\bf claim} goes as follows.\\

One has exact  \M  {\bf \IF } $$\forall\ket{\phi}_A^\mathbf{i},\enskip\forall k:\enskip\Big(F_B^kU_{AB}\Big)\Big(\ket{\phi}_A^\mathbf{i}
\ket{\phi}_B^\mathbf{i}\Big)= \Big(U_{AB}E_A^k\Big)\Big(\ket{\phi}_A^\mathbf{i}
\ket{\phi}_B^\mathbf{i}\Big) \eqno{(B.1)}$$ is valid.\\

One {\it proves necessity} as follows.
The completeness relation \$\sum_{k'}E_A^{k'}=I_A\$, use of the \CC (5), and orthogonality and idempotency of the \$F_B^k\$ projectors enable one to write for each \$k\$ value : $$F_B^kU_{AB}\ket{\phi}_A^\mathbf{i}
\ket{\phi}_B^\mathbf{i}=$$ $$\sum_{k'}||E_A^{k'}\ket{\phi}_A^\mathbf{i}||\times F_B^k
U_{AB}\Big( E_A^{k'}\ket{\phi}_A^\mathbf{i}\Big/ ||E_A^{k'}\ket{\phi}_A^\mathbf{i}||\Big)
\ket{\phi}_B^\mathbf{i}=$$ $$\sum_{k'}||E_A^{k'}\ket{\phi}_A^\mathbf{i}||\times F_B^k
\mathbf{F_B^{k'}}U_{AB}\Big( E_A^{k'}\ket{\phi}_A^\mathbf{i}\Big/ ||E_A^{k'}\ket{\phi}_A^\mathbf{i}||\Big)
\ket{\phi}_B^\mathbf{i}=$$ $$||E_A^k\ket{\phi}_A^\mathbf{i}||\times F_B^kU_{AB}\Big( E_A^k\ket{\phi}_A^\mathbf{i}\Big/ ||E_A^k\ket{\phi}_A^\mathbf{i}||\Big)
\ket{\phi}_B^\mathbf{i}.$$ Thus,
$$F_B^kU_{AB}\ket{\phi}_A^\mathbf{i}
\ket{\phi}_B^\mathbf{i}=||E_A^k\ket{\phi}_A^\mathbf{i}||\times F_B^kU_{AB}\Big( E_A^k\ket{\phi}_A^\mathbf{i}\Big/ ||E_A^k\ket{\phi}_A^\mathbf{i}||\Big)
\ket{\phi}_B^\mathbf{i}.\eqno{(B.2)}$$

Finally, on account of (5) again, we can omit \$F_B^k\$, so that, after cancelation, one obtains: $$F_B^kU_{AB}\ket{\phi}_A^\mathbf{i}
\ket{\phi}_B^\mathbf{i}=
U_{AB}E_A^k\ket{\phi}_A^\mathbf{i}
\ket{\phi}_B^\mathbf{i}.$$ 

The cancellation cannot be done if \$||E_A^k\ket{\phi}_A^\mathbf{i}|=0\$. But the claimed relation (B.1) is still valid becauae the RHS is obviously zero, and so is the LHS as seen in (B.2).\\

To {\it prove sufficiency}, let $$\Big( U_{AB}E_A^k\Big) \Big(\ket{\phi}_A^\mathbf{i}\ket{\phi}_B^\mathbf{i}\Big)= \Big(F_B^kU_{AB}\Big)
\Big(\ket{\phi}_A^\mathbf{i}\ket{\phi}_B^\mathbf{i}\Big)$$ be valid for all \$k\$ values, and let \$\ket{\phi}_A^\mathbf{i}=E_A^{\bar k}\ket{\phi}_A^\mathbf{i}\$ be satisfied for a fixed value  \$k\equiv \bar k\$. Then, one has in particular $$\Big(U_{AB}E_A^{\bar k}\Big) \Big(\ket{\phi}_A^\mathbf{i}\ket{\phi}_B^\mathbf{i}\Big)= \Big(F_B^{\bar k}U_{AB}\Big)
\Big(\ket{\phi}_A^\mathbf{i}\ket{\phi}_B^\mathbf{i}\Big).$$ One can here omit \$E_A^{\bar k}\$ due to the assumed definite value in \$\ket{\phi}_A^\mathbf{i}\$ (cf (5)), and thus the explicit form of the \CC (5) is obtained. {\it This ends the proof.}\\

{\bf Appendix C. Proof of the Probability Reproducibility Condition }\\
\noindent
The \PRC reads:
$$\forall\ket{\phi}_A^i,\enskip\forall k:\quad\bra{\Phi}_{AB}^f F_B^k
\ket{\Phi}_{AB}^f=\bra{\phi}_A^i
E_A^k\ket{\phi}_A^i,\eqno{(C.1)}$$

{\it Proof.} Utilizing definition (3), the dynamical condition (6), and the idempotency of \$F_B^k\$ and of \$E_A^k\$, one can see that  $$LHS(C.1)=\bra{\phi}_A^i\bra{\phi}_B^i
\Big(E_A^kU_{AB}^{\dag}\Big) \Big({U_{AB}E_A^k}\Big)\ket{\phi}_A^i
\ket{\phi}_B^i=RHS(13).$$\\


\begin{thebibliography}{99}

\bibitem{FHenvar2}
F. Herbut, {\it Eur. Phys. J. Plus}  {\bf 127}, 14 (2012).

\bibitem{FHijqi1}
F. Herbut, {\it Int. J. Quant. Inf.} {\bf 12},   1450032 (16 pages) (2014).

\bibitem{FHarxiv}
F. Herbut, Arxiv:1412.7862 (2014).

\bibitem{BLM}
P. Busch, P. K. Lahti, and P. Mittelstaedt, {\it The Quantum Theory of Measurement}, 2nd edition (Springer, Berlin, 1996).

\bibitem{Peres}
A. Peres, {\it Quantum Theory: Concepts and Methods}, (Kluwer Ac. Publ., Dordrecht, 1993).

\bibitem{Bell}
J. Bell, {\it Physics World} August,  33 (1990).

\bibitem{FHOverMeas}
F. Herbut, Arxiv:1511.07402 (2015).

\bibitem{Farhi}
E. Farhi and J. GOLDSTONE, {\it
Ann. Phys. (N. Y.)} {\bf 192}, 368(1989).


\end{thebibliography}
\end{document}